\documentclass[%
reprint,
superscriptaddress,
showpacs,
showkeys,
amsmath,amssymb,amsfonts,
aps,
pra,
floatfix,
a4paper,
twocolumn
]{revtex4-2}
\usepackage{shellesc}
\usepackage{iftex}
\ifluatex 	
	\usepackage{fontspec}	
	\defaultfontfeatures{Ligatures=TeX}
\else
	\usepackage[utf8]{inputenc}
	\usepackage[T1]{fontenc}
\fi
\usepackage{graphicx}
\usepackage[usenames,dvipsnames]{xcolor}
\usepackage{dcolumn}
\usepackage{bm}
\usepackage{bookmark}	
\usepackage[all]{hypcap}

\makeatletter			
\newcommand*\nobreakhyphen{\hbox{-}\nobreak\hskip\z@skip}
\makeatother			
\usepackage{ulem}
\newcommand{\XC}[1]{\textcolor{black}{#1}}
\begin{document}
\title{Robust quantum control by smooth quasi-square pulses}
\author{Jing-jun Zhu}
\affiliation{Laboratoire Interdisciplinaire Carnot de Bourgogne, CNRS UMR 6303, Universit\'e de Bourgogne Franche-Comt\'e, BP 47870, F-21078 Dijon, France}%
\author{Xavier Laforgue}
\affiliation{Laboratoire Interdisciplinaire Carnot de Bourgogne, CNRS UMR 6303, Universit\'e de Bourgogne Franche-Comt\'e, BP 47870, F-21078 Dijon, France}%
\author{Xi Chen}
\affiliation{Department of Physical Chemistry, University of the Basque Country UPV/EHU, Apartado 644, 48080 Bilbao, Spain}
\author{St\'ephane Gu\'erin}
\email{sguerin@u-bourgogne.fr}
\affiliation{Laboratoire Interdisciplinaire Carnot de Bourgogne, CNRS UMR 6303, Universit\'e de Bourgogne Franche-Comt\'e, BP 47870, F-21078 Dijon, France}%
\date{\today}
\begin{abstract}
Robust time-optimal control is known to feature constant (square) pulses.  We analyze fast adiabatic dynamics that preserve robustness by using alternative smooth quasi-square pulses, typically represented by hyper-Gaussian pulses. We here consider the two protocols, robust inverse optimization and time-contracted adiabatic passage, allowing the design of the same pulse shape in both cases. The dynamics and their performance are compared. The superiority of the former protocol is shown.
\end{abstract}
\maketitle
\section{Introduction}
Robust and high-fidelity control is a central prerequisite for quantum information processing \cite{TrainingSC}. The problem has been addressed differently by adiabatic techniques and composite pulses. The former is based on the adiabatic theorem, which mathematically requires an infinitely slow dynamics or infinitely large coupling strengths. 
In practical applications, necessitating fast dynamics and low coupling, adiabatic passage is applied with finite times and is only approximate.
Methods to accelerate it and to make it accurate while preserving its relative robustness have been proposed.
One can mention (i) parallel adiabatic passage (PAP) \cite{parallel1,parallel2,parallel3,parallel4}, which allows the design of the shape of the control parameters minimizing the non-adiabatic transfer; (ii) counterdiabatic driving \cite{CD1,CD2,CD3} (included into a more general class of techniques usually referred to as shortcut to adiabaticity (STA) \cite{STA}), which is based on the addition of a coupling term that compensates the non-adiabatic coupling, and (iii)  local optimization \cite{localadiabatic}, referred to as FAQUAD \cite{FAQUAD}, based on the control of the dynamics via a single parameter through the gap of an avoided crossing. Composite pulses, on the other hand, use a sequence of precise $\pi$-pulse transfers and an error resilience by (static) phases, see e.g. \cite{Vitanov}. More general time-dependent phases can be used, as in the single-shot shaped pulse (SSSP) method, \cite{SSSP,Hybrid}, or more generally STA based on reverse engineering techniques \cite{STA}, which offer faster dynamical routes featuring both robustness and high-fidelity.


Methods to determine the fastest (or least energetic) processes under given conditions can be determined by optimal control, 
through the minimization of a cost functional, such as, e.g., Pontryagin maximum principle \cite{Pontryagin}, which provides a global (absolute) optimum for simple systems. Numerical optimization, such as, e.g., gradient-based optimization algorithm (GRAPE algorithm \cite{GRAPE}), provide only local optimal solutions. 
Only recently, techniques merging robustness and optimality have been proposed \cite{Extended,Barnes}, including a method based on inverse optimization (RIO) \cite{Dridi,Xavier}. In particular, time-optimal robust processes have been shown to feature driving pulses of constant amplitude, i.e. square pulses. However such square wave forms are difficult to generate in practice and can produce inaccuracies at the discontinuities. It is thus desirable to develop techniques that can systematically produce smooth control pulses close to ideal optimal square pulses.

In this paper, we investigate the construction of robust quantum control using smooth quasi-square pulses in the two regimes, optimal and adiabatic: (i) robust optimization \cite{Dridi,Xavier} and (ii) rescaled adiabatic passage with a contracted time \cite{TR}. 
In both regimes, we use a formulation allowing a design of the pulse approaching square pulses. We choose a hyper-Gaussian shape (of high order, i.e. featuring fast ramps) as a smooth quasi-square pulse approaching the ideal time-optimal square pulse.

We compare the two protocols in terms of dynamics and robustness. 
It is shown these two regimes feature very different dynamical process and that, for an identical pulse area, RIO offers a more efficient robustness compared to a time-contracted adiabatic passage, where adiabaticity is merely approximate. By increasing the pulse area over 50$\%$, the latter can achieve performance comparable to the RIO  protocol.

The paper is organized as follows. Section 2 introduces the model and the methods offering fast and close to time-optimal robust processes, namely RIO and time-contracted adiabatic passage. Section 3 presents the results with the use of a hyper-Gaussian pulse as a quasi-square pulse. Other quasi-square pulses are investigated for time-contracted adiabatic passage made on an expansion combining a linear and sine terms of the rescaled function. Section 4 concludes.
\section{Model and methods}
We consider the model 
\begin{equation}
	H_0=\frac{\hbar}{2}\left[
	\begin{array}{cc}
		-\Delta & \Omega\\
		\Omega & \Delta
	\end{array}
	\right]
	\label{H},
\end{equation}
corresponding to a two-level system driven by a quasi-resonant pulse in rotate-wave approximation,
where $\Omega$ and $\Delta=\omega_{2}-\omega_{1}-\omega_{0}-\dot{\phi}$ denote the (time-dependent) Rabi frequency and detuning, respectively. Here $\omega_{21}=\omega_{2}-\omega_{1}$ is frequency difference between the two levels and $\omega_{0}+\dot{\phi}$ is the time-dependent instantaneous frequency of the laser field (with $\omega_0$ its mean frequency). Under resonance condition $\omega_0=\omega_2-\omega_1$, the phase of the field reads: $\phi(t)=\int_{t_i}^{t}\Delta(s)ds$.  

The solution of the time dependent Schr\"odinger equation (TDSE)
$i\hbar\frac{\partial}{\partial t}\vert\phi_0(t)\rangle=H_0\vert\phi_0(t)\rangle$ is conveniently
parameterized with three angles: the mixing angle
$\theta\equiv\theta(t)\in[0,\pi]$, the internal (or relative) phase $\varphi\equiv\varphi(t)\in[-\pi,\pi]$
and a global phase $\gamma\equiv\gamma(t)\in[0,2\pi]$ as
\begin{equation}
	\label{phizero}
	\vert\phi_0(t)\rangle=\left[\begin{matrix}
		e^{i\varphi/2}\cos(\theta/2)\\
		e^{-i\varphi/2}\sin(\theta/2)
	\end{matrix}\right] e^{-i\gamma/2}.
\end{equation}
Inserting it into the TDSE, we obtain
\begin{subequations}
	\label{dotpara}
\begin{align}
		\label{dottheta}
		\dot\theta&=\Omega\sin\varphi,\\
		\label{dotphi}
		\dot\varphi&=\Delta+\Omega\cos\varphi\cot\theta,\\
		\label{dotgamma}
		\dot\gamma&
		=\Omega \frac{\cos\varphi}{\sin\theta}=\dot\theta \frac{\cot\varphi}{\sin\theta},
\end{align}
\end{subequations}
where the dot represents the derivation with respect to time $t$.

One can inversely determine the controls from the dynamical angles:
\begin{subequations}
\begin{eqnarray}
        \label{delta}
		&\Delta=\dot{\varphi}-\dot{\gamma}\cos\theta,\\
		&\Omega=\sqrt{\dot{\theta}^{2}+\dot{\gamma}^{2}\sin^{2}\theta}=|\dot{\gamma}|\sqrt{\bigl(\dot{\widetilde{\theta}}\bigr)^{2}+\sin^{2}\tilde{\theta}},
		\label{omega}
\end{eqnarray}
\end{subequations}
with $\dot{\tilde{\theta}}\equiv d\tilde{\theta}/d\gamma$, where $\theta$ as a function of $\gamma$ is denoted as $\widetilde\theta(\gamma)$.

\subsection{Robust inverse optimization}
To take into account robustness, we consider the perturbed Hamiltonian of the form $H_{\lambda}=H_0+\lambda V$, 
\begin{equation}
	\label{Halpdelt}
	H_{\alpha,\beta,\delta}=\frac{\hbar}{2}\left[\begin{array}{cc} -\Delta & \Omega\\ \Omega & \Delta\end{array}\right]
	+\frac{\hbar}{2}\left[\begin{array}{cc} -\delta & \alpha\Omega + \beta \\ \alpha\Omega + \beta & \delta\end{array}\right],
\end{equation}
where $\alpha$ is a coefficient modifying the Rabi field amplitude (pulse inhomogeneities), $\delta$ features inhomogeneous broadening or a slow stochastic noise in the energy levels of the qubit (i.e. considered in a quasi-static representation), and $\beta$ (considered real) a slow stochastic transverse noise.
We consider the family of the state solutions $\vert\phi_{0}(t_f)\rangle$ 
that reach exactly the target state $|\phi_T\rangle$ at the end of the process, at $t=t_f$, without perturbation, from a given initial state at $t=t_i$. Among these solutions, we look for those that reach it in a robust way as defined below.

The SSSP method \cite{SSSP} consists in expanding the projection of the given state solution $\vert\phi_{\lambda}(t_f)\rangle$ on the given target state $|\phi_T\rangle$ at the end of the process with respect to $\lambda$, giving the probability amplitude:
\begin{equation}
	\label{expansion}
\langle\phi_T|\phi_{\lambda}(t_f)\rangle=1+O_1+O_2+O_3+\cdots,
\end{equation}
where $O_n$ denotes the term of total order $n$: $O_n\equiv O(\lambda^n)$ 
\begin{align}
	\label{Errors}
		O_1&=-i\int_{t_i}^{t_f}\langle\phi_0(t)|V(t)|\phi_0(t)\rangle dt\equiv
		-i\int_{t_i}^{t_f}e(t)dt,\\
		O_2&=(-i)^2\int_{t_i}^{t_f}dt\int_{t_i}^{t}dt'[e(t)e(t')
		+f(t)\bar f(t')],
		\label{ErrorsO2}
\end{align}
with 
\begin{align}
		e
		&=-\frac{1}{2}(\delta\cos\theta-\alpha\dot\gamma\sin^2\theta - \beta\sin\theta\cos\varphi),\\
		f
		&=\frac{1}{2}\Bigl[\delta\sin\theta
		+\alpha\Bigl(\frac{1}{2}\dot\gamma\sin2\theta-i\dot\theta\Bigr)
		\nonumber\\
		&\qquad\qquad{}+\beta(\cos\varphi\cos\theta-i\sin\varphi)\Bigr]e^{i\gamma}.
\end{align}
We will consider in this paper robustness with respect to pulse inhomogeneities ($\alpha$-robustness), i.e. $\delta=\beta=0$, $\lambda\equiv\alpha$, for population transfer at second order: $P_T(t_f)=1-(O_2+\bar O_2+\bar O_1O_1)=1-\vert \int_{t_i}^{t_f} f(t) dt\vert ^2$.
Robustness up to second order corresponds thus to the two constraints 
\begin{eqnarray}
	\label{constraints}
	\int_{t_i}^{t_f} {\cal R}[f(t)] dt=0,\qquad \int_{t_i}^{t_f} {\cal I}[f(t)] dt=0.
\end{eqnarray}
Optimal trajectory of the dynamical angles with respect to a given cost, \textit{e.g.} the pulse area $\int_{t_i}^{t_f} \vert \Omega (t)\vert dt $, the pulse energy $\int_{t_i}^{t_f} \vert \Omega (t)\vert^2 dt $, or the pulse duration (time-optimization), that satisfy the constraints \eqref{constraints} can be obtained via an Euler-Lagrange formulation, and the controls are determined from this trajectory via \eqref{delta} and \eqref{omega}. This protocol has been named robust inverse optimization (RIO). 

\begin{figure}[t]
	\begin{center}
		\includegraphics[scale=0.75]{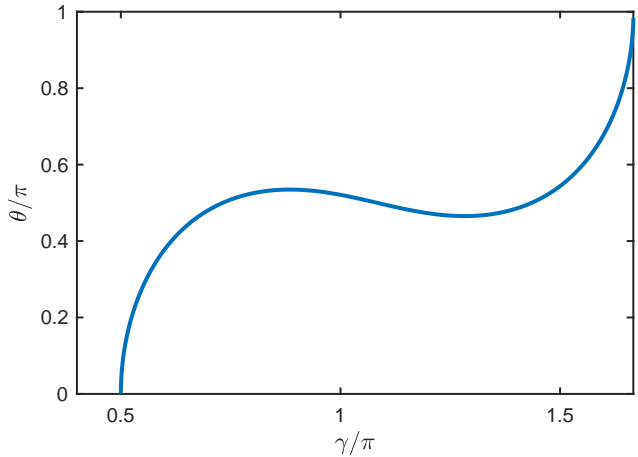}
		\caption{Optimal $\alpha$-robust geodesic $\widetilde\theta(\gamma)$ in the dynamical variable space $(\gamma,\theta)$ leading to complete population transfer ($\theta_f=\pi$) from the ground state ($\theta_i=0$) (leading to $\gamma_f=5\pi/3$ and $\varphi_f=\pi/2$).}
		\label{figTraject}
	\end{center}
\end{figure}

We can analyze the problem of complete or partial population transfer from the ground state ($\theta(t_i)\equiv \theta_i=0$, $\gamma(t_i)\equiv \gamma_i=\varphi(t_i)\equiv \varphi_i=\pi/2$), 
which is optimal with respect to the pulse area, by fixing the final targeted mixing angle $\theta(t_f)\equiv \theta_f=\theta_0$ and internal phase $\varphi (t_f)\equiv\varphi_f=\varphi_0$ (the final value of $\gamma$ being irrelevant for population transfer).
In this paper, we consider the complete population transfer problem, i.e. $\theta_f=\pi$. 
It corresponds to the trajectory of $\theta$ as a function of $\gamma$, denoted as $\widetilde\theta(\gamma)$, 
displayed in Fig. \ref{figTraject}, as determined in \cite{Dridi}.
Time-optimal robust trajectory (for a given peak amplitude $\Omega_0$) corresponds to a special time-parametrization of the dynamical angle $\gamma(t)$ given by the constant of motion \begin{eqnarray}
	\dot\theta^2+\dot\gamma^2\sin^2\theta=\Omega_0^2,
\end{eqnarray}
with a constant pulse $\Omega=\Omega_0$, i.e. a square pulse, of duration $T_{\min}=t_f-t_i$:
\begin{eqnarray}
	\label{Tmin}
	T_{\min}=\frac{1}{\Omega_0}\int_{\gamma_i}^{\gamma_f}d\gamma\,\sqrt{\bigl(\dot{\widetilde\theta}\bigr)^2+\sin^2\widetilde\theta}\approx 5.84/\Omega_0.
\end{eqnarray}

%
The trajectory in Fig. \ref{figTraject} offers thus infinitely many optimal solutions  (robust with respect to the pulse area) depending on the time parametrization of $\gamma(t)$. In order to approach the ideal time-optimal robust dynamics, we impose a parametrization given by a smooth quasi-square pulse of same peak amplitude and same area as that of the square pulse. This is modeled by a hyper-Gaussian pulse of high-order even $n$ and width $\sigma$:
\begin{eqnarray}
	\Omega(t)=\Omega_{0}{\rm{exp}}(-(t/\sigma)^{n}),
\end{eqnarray}
such that $\int \Omega(t) dt=2\Omega_0\sigma \Gamma(1/n)/n=\Omega_0 T_{\min}$, i.e. 
\begin{eqnarray}
	\sigma=\frac{n T_{\min}}{2\Gamma(1/n)}.
\end{eqnarray}
One could choose in principle any even parameter $n$; we expect a longer process for a smaller $n$ deviating more from the ideal (optimized) squared pulse.

The parametrization of $\gamma(t)$ is more precisely determined from the identity of the partial pulse areas:
\begin{eqnarray}
	\Omega_0\int_{t_{i}}^{t}{\rm {exp}} (-(t/\sigma)^{n})dt =\int_{\gamma_i}^{\gamma(t)}d\gamma\,\sqrt{\bigl(\dot{\widetilde\theta}\bigr)^2+\sin^2\widetilde\theta}.
\end{eqnarray}

\subsection{Time-contracted adiabatic passage}
Adiabatic passage means a dynamics which follows approximately the instantaneous eigenvector continuously connected to the initial condition (when $\Omega=0$).
The eigenvectors read
\begin{eqnarray}
		&\vert \psi_{+}\rangle=\cos(\Theta/2)|1\rangle + \sin(\Theta/2)|2\rangle,\\
		&\vert \psi_{-}\rangle= \cos(\Theta/2)|2\rangle-\sin(\Theta/2)|1\rangle,
\end{eqnarray}
with the mixing angle $\Theta=\arctan(-\Omega/\Delta)$, $0\le\Theta\le\pi$, and the corresponding eigenenergy $\lambda_{\pm}=\pm\hbar\Omega_{a}/2$ with $\Omega_{a}=\sqrt{\Delta^{2}+\Omega^{2}}$ .

To achieve an efficient and robust population transfer by adiabatic passage, we consider the parallel adiabatic passage technique \cite{parallel1}, where we define an analytic shape $\Lambda(t)$ of the Rabi frequency: $\Omega(t)=\widetilde\Omega_0\Lambda(t)$, from which we design the detuning $\Delta(t)$ such that the distance between the eigenvalues remain constant at all times:
\begin{align}
	\label{pulses}
		\Omega(t)&=\widetilde\Omega_{0}\Lambda(t/T),\\
		\Delta(t)&=\widetilde\Omega_{0} \,{\rm {sign}}(t/T) \sqrt{1-\Lambda^2(t/T)},
\end{align}
giving $\Omega_a=\widetilde\Omega_0$.
The parameter $T$ characterizes the width of the pulse. 

Renormalizing the time $s=t/T$ in the Schr\"odinger equation 
shows that the dynamics in the adiabatic basis $\vert\psi_a\rangle=T_a^{-1}\vert\psi\rangle$ with $T_a=[\vert\psi_-\rangle,\vert\psi_+\rangle]$, $D_a=T_a^{\dagger}HT_a={\rm{diag}}(\lambda_-,\lambda_+)$, given by
\begin{eqnarray}
	&i\hbar\frac{\partial}{\partial s}\vert\psi_a\rangle=T\widetilde\Omega_0\left[\begin{array}{cc}
		-1 & 0\\
		0 & 1 \end{array}
	\right]\vert\psi_a\rangle -i\hbar T_a^{\dagger} \dot T_a \vert\psi_a\rangle,
\end{eqnarray}
is better approximated by the diagonal Hamiltonian $D_a$ when the dimensionless quantity characterizing the pulse area is large $T\widetilde\Omega_0\gg 1$. It also shows that accelerating the dynamics with a shorter time $T'<T$ gives trivially the same (adiabatic) dynamics but for a larger peak Rabi frequency $\widetilde\Omega'_0=\widetilde\Omega_0 T/T'$ (with the same pulse area).

The method proposed in \cite{TR} shows a method to define, from a given adiabatic dynamics, a rescaled adiabatic passage featuring the same transfer and the same robustness as the original one. We will apply it with a time contraction and will refer it to as time-contracted adiabatic passage (TCAP). It consists in defining a rescaled time $\tau$, defined by the rescaled function $g(\tau)$ such that $t=g(\tau)$, in which the dynamics realizes exactly the same final state but in a shorter time. This is achieved when $\tau_f-\tau_i < t_f-t_i$, 
and $\tau_i=g^{-1}(t_i)$ and $\tau_f=g^{-1}(t_f)$ the initial and final times in the new frame, respectively, such that the state functions are (i) the same ones at initial time, (ii) the same ones at final times, and that the Hamiltonians are (iii) the same ones initially and (iv) the same ones finally. More explicitly, the rescaled dynamics corresponds to the Schr\"odinger equation
\begin{eqnarray}
	i \hbar \frac{\partial}{\partial \tau} \psi_r(\tau)= H_r(\tau) \psi_r(\tau), \ H_r(\tau)=g'(\tau) H(g(\tau)),
\end{eqnarray}
with the rescaled state and Hamiltonian, respectively,
\begin{eqnarray}
	\psi_r(\tau)=\psi(g(\tau)),\quad H_r(\tau)=g'(\tau) H(g(\tau)).
\end{eqnarray}
The above conditions are satisfied when
\begin{eqnarray}
	\label{def_TR}
	g_i\equiv g(\tau_i)=t_i,\ g(\tau_f)=t_f,\ g'(\tau_i)= g'(\tau_f)=1.
\end{eqnarray}
The corresponding rescaled controls can be then expressed as
\begin{align}
		\label{pulse_TR}
		\Omega_c(\tau)&=\widetilde\Omega_{0}g^{\prime}(\tau)\Lambda(g(\tau)/T),\\
		\label{det_TR}
		\Delta_c(\tau)&=\widetilde\Omega_{0}g^{\prime}(\tau)  \,{\rm{sign}}(g(\tau)/T) \sqrt{1-\Lambda^2(g(\tau)/T)}.
\end{align}
We can notice that the pulse area is invariant: $\int \Omega_r(\tau) d\tau =\widetilde\Omega_{0} \int \Lambda(g/T) dg  =\int \Omega(t)dt$.

We consider the time contraction defined as 
\begin{eqnarray}
	\label{contraction}
	\tau_i=t_i/a, \quad \tau_f=t_f/a.
\end{eqnarray}
which contracts the duration of the process by a factor $a>1$.
Our goal is to design, for a given time-contraction $a$, the rescaled function $g(t)$ such that the pulse features a quasi-square shape.


We search for quasi-square pulse following two strategies: (i) we start from an expansion of $g$ with free coefficients satisfying \eqref{def_TR} and \eqref{contraction},  and determine the optimal coefficients that minimize the pulse amplitude \eqref{pulse_TR}, (ii) we impose the rescaled pulse \eqref{pulse_TR} to be an hyper Gaussian pulse (of high order). In both cases, we infer the corresponding detuning \eqref{det_TR}. 

We remark that we focus in this paper more specifically on the parallel adiabatic passage with a Gaussian pulse $\Lambda(t)=e^{-t^2}$, but that any other adiabatic passage design can be used.

\section{Robust control with quasi-square pulse}
\subsection{Robust inverse optimization with hyper-Gaussian pulse}

\begin{figure}
	\begin{center}
		\includegraphics[scale=0.75]{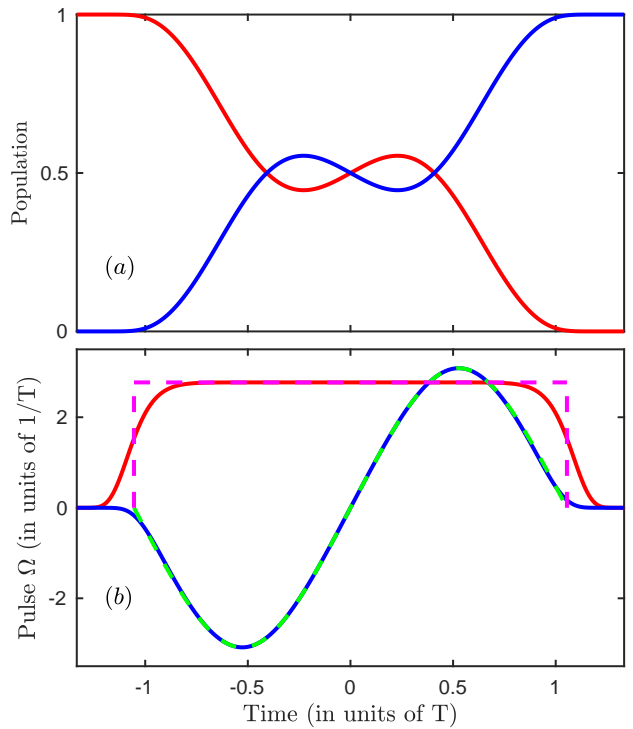}
		\caption{(a) The hG-RIO population dynamics (ground state: red line, excited state: blue line). (b) The quasi-square hyper-Gaussian pulse (solid lines) and  the time-optimal ideal square pulse (dashed line) with their corresponding oscillating detunings (almost undistinguishable at the scale of the figure, except at the beginning and at the end). 
			} 
		\label{fig2}
	\end{center}
\end{figure}

Fig. \ref{fig2} presents the dynamics driven by RIO with a hyper-Gaussian pulse (of high order $n=14$ and $\sigma\approx 1.095T$, 
corresponding to the same pulse as in the TCAP, see  Fig. \ref{figAdiabPulses}, i.e. $\Omega_0\approx 2.77/T$) replacing the ideal optimal square pulse (of same area 5.84). It will be referred in short to as hyper-Gaussian RIO (hG-RIO).  As expected the corresponding detunings are very similar.
The detuning associated to the hyper-Gaussian pulse also appears smooth (i.e. regularized at the beginning and at the end of the process).

\subsection{Time-contracted adiabatic passage with linear-sine expansion}
\begin{figure}
	\begin{center}
		\includegraphics[scale=0.73]{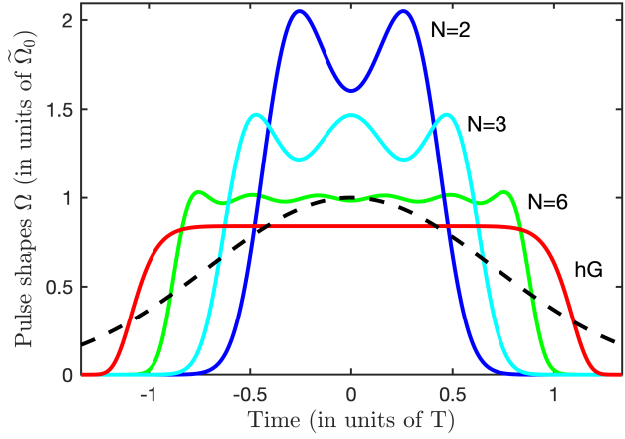}
		\caption{Time-contracted pulse shapes for $N=2,3,6$ where the coefficients $C_n$'s have been optimized to minimize the peak of the pulse and the hyper-Gaussian pulse for $n=14$ (labelled as hG) from the original Gaussian pulse (dashed line) that has been time-contracted.} 
		\label{figAdiabPulses}
	\end{center}
\end{figure}

We analyze TCAP using a $g$ function combing a linear term and a sine expansion with $N$ terms:
\begin{equation}
	g(\tau)=
	a\tau+\sum_{n=1}^N C_{n}\sin\left[\frac{2n\pi a}{t_{f}-t_{i}}\left(\tau-\frac{t_{i}}{a}\right)\right].
	\label{f0}
\end{equation}
This ensures in a simple way that \eqref{def_TR} with \eqref{contraction} are satisfied, while offering free parameters to design the shape of the pulse \eqref{pulse_TR}. More specifically, $g(t_i/a)=t_i$ and  $g(t_f/a)=t_f$ are satisfied by the linear term $a\tau$. 
The derivatives give
\begin{equation}
	g'(t_i/a)=g'(t_f/a)=
	a+ \frac{2\pi a}{t_{f}-t_{i}} \sum_{n=1}^N n C_{n},
\end{equation}
which imposes a condition on the sum
\begin{equation}
	\label{constrCn}
	\sum_{n=1}^N nC_{n}=\frac{(1-a)  (t_{f}-t_{i})}{2\pi a}.
\end{equation}
This means that, for a given $N$, we can choose freely $N-1$ coefficients and the last one is constrained by \eqref{constrCn}. 
We follow the protocol: starting with a Gaussian pulse, we determine numerically for a given $N$ the coefficients $C_{n<N}$ with \eqref{constrCn}, leading to the smallest pulse peak amplitude.
We consider the expansion \eqref{f0} of $g$ for various $N$. 
Figure \ref{figAdiabPulses} shows the resulting shape of the pulses for $a=3$. We can notice as expected that, for larger $N$, the pulse tends to a quasi-square pulse with a decreasing peak amplitude. 

\subsection{Time-contracted adiabatic passage with hyper-Gaussian pulse}
In this section, we impose the time-contracted pulse to be a hyper-Gaussian pulse of high-order (even) $n$:
\begin{eqnarray}
	\label{pulse_TR_hG}
	\Omega_c(\tau)=\widetilde\Omega_{0}g^{\prime}(\tau)\Lambda(g(\tau)/T)
	\equiv \widehat\Omega_0\exp\bigl(-(\tau/\sigma)^n\bigr),
\end{eqnarray}
starting from a Gaussian pulse: $\Lambda(t)=e^{-t^2}$. It is referred to as hG-TCAP.
We derive the corresponding $g(t)$ function by integration.
We obtain (using $\tau_i=-\infty$ and the condition $g_i=t_i=-\infty$) for $\tau\ge0$
\begin{eqnarray}
	{\rm{erf}}\biggl(\frac{g}{T}\biggr)= \frac{2\sigma \widehat\Omega_{0}}{\sqrt{\pi}T\widetilde\Omega_{0}n}\biggl[ 2\Gamma\biggl(\frac{1}{n}\biggr)- \Gamma\biggl(\frac{1}{n},\frac{\tau^n}{\sigma^n}\biggr)\biggr]-1,
\end{eqnarray}
and for $\tau\le0$
\begin{eqnarray}
	{\rm{erf}}\biggl(\frac{g}{T}\biggr)=\frac{2\sigma \widehat\Omega_{0}}{\sqrt{\pi}T\widetilde\Omega_{0}n}\Gamma\biggl(\frac{1}{n},\frac{\tau^n}{\sigma^n}\biggr)-1.
\end{eqnarray}
Using an antisymmetric $g$, we have $g(0)=0$, i.e.
\begin{eqnarray}
	\sigma \widehat\Omega_{0}=\frac{\sqrt{\pi}T\widetilde\Omega_{0}n}{2 \Gamma\bigl(\frac{1}{n}\bigr)},
\end{eqnarray}
leading to
\begin{align}
	{\rm{erf}}\biggl(\frac{g}{T}\biggr)&= \frac{1}{\Gamma\bigl(\frac{1}{n}\bigr)}\biggl[ 2\Gamma\biggl(\frac{1}{n}\biggr)- \Gamma\biggl(\frac{1}{n},\frac{\tau^n}{\sigma^n}\biggr)\biggr]-1,\quad \tau\ge0,\nonumber\\
	{\rm{erf}}\biggl(\frac{g}{T}\biggr)&=\frac{1}{\Gamma\bigl(\frac{1}{n}\bigr)}\Gamma\biggl(\frac{1}{n},\frac{\tau^n}{\sigma^n}\biggr)-1,\quad \tau\le0.
\end{align}
From the derivative evaluated at $\tau_f=t_f/a$ for which $g(\tau_f)=t_f$ and $g'(\tau_f)=1$, we obtain
\begin{eqnarray}
	1=\frac{\sqrt{\pi}Tn}{2\sigma \Gamma\bigl(\frac{1}{n}\bigr)} \frac{e^{-(t_f/(a\sigma))^n}}{e^{-(t_f/T)^2}} ,
\end{eqnarray}
i.e.
\begin{eqnarray}
	\sigma = \frac{t_f}{a[\ln(\widehat\Omega_{0}/\widetilde\Omega_{0})+(t_f/T)^2 ]^{1/n}}\to  \frac{t_f}{a} (T/t_f)^{2/n}.
\end{eqnarray}
This shows that the width $\sigma$ tends to infinity for an infinite $t_f$, with an amplitude $\widehat\Omega_{0}$ going to 0.
In practice, we have to fix $t_f$ (and symmetrically $t_i$) to a finite value such that $\Lambda(t_f)=\epsilon$ is negligibly small.

\begin{figure}
	\begin{center}
		\includegraphics[scale=0.75]{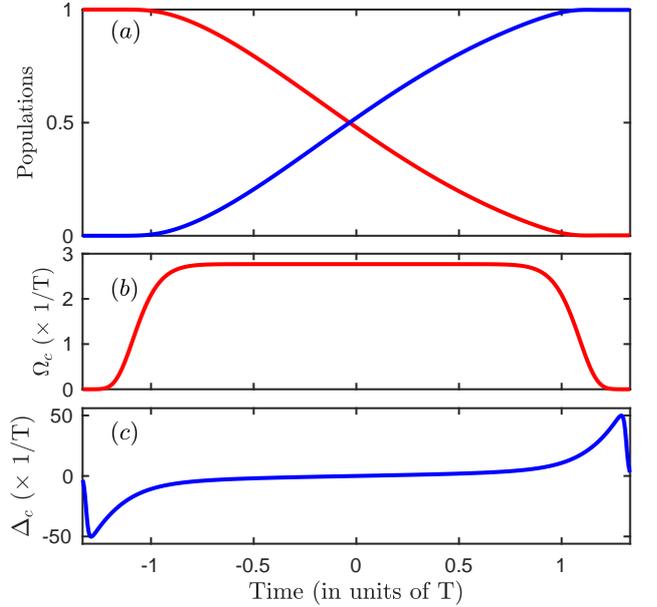}
		\caption{(a) The hG-TCAP population dynamics (ground state: red line, excited state: blue line) induced by the hyper-Gaussian pulse (b) (taken from Fig. \ref{figAdiabPulses}) and the corresponding detuning (c).} 
		\label{fig4}
	\end{center}
\end{figure}

Figure \ref{figAdiabPulses} shows the resulting hyper-Gaussian pulse shape for $n=14$ generated from $a=3$ and $t_f=4T$, i.e. $\widehat\Omega_{0}\approx 0.84\widetilde\Omega_0=\Omega_0=2.77/T$ and $\sigma\approx 1.095T$.
Figure \ref{fig4} shows the hG-TCAP dynamics leading to the population transfer resulting from the hyper-Gaussian pulse and the detuning [determined from \eqref{det_TR}]. The latter features early and late high-amplitude bumps inducing far-from-resonance dynamics, which allows the fast ramps of the hyper-Gaussian pulse to satisfy adiabaticity. We highlight that these bumps result directly from the time-contraction of the adiabatic process. Between the two bumps, the detuning shows a slow quasi-linear chirp crossing the resonance during the quasi-plateau of the hyper-Gaussian pulse, which ensures adiabatic transfer.

\begin{figure}
	\begin{center}
		\includegraphics[scale=0.75]{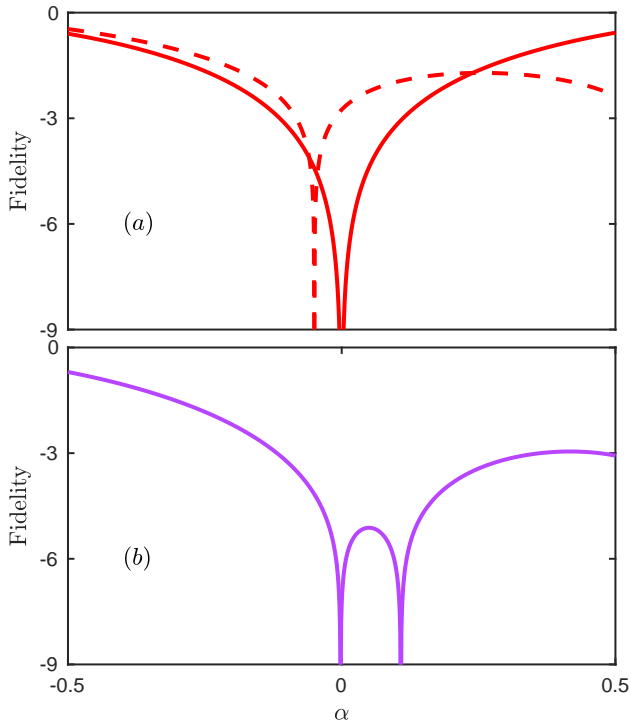}
		\caption{(a) Fidelity (dimensionless) of the transfer (in a base 10 logarithmic scale) as a function of the relative deviation $\alpha$ in the pulse amplitude $\Omega_0(1+\alpha)$ for hG-RIO (solid red line) and hG-TCAP (dashed red line) with the same hyper-Gaussian pulse shape and area (5.84). (b) Fidelity of the hG-TCAP transfer (in a base 10 logarithmic scale) with a pulse area more than 50\% larger (9.22) that leads to a fidelity comparable to that of hG-RIO (upper frame).} 
		\label{fig5}
	\end{center}
\end{figure}
\subsection{Robustness}

Figure \ref{fig5}, upper frame, compares the robustness profile against a relative deviation of the pulse amplitude for hG-RIO and hG-TCAP, respectively, using the same pulse, via the fidelity of the transfer ${\cal F}=1-P$, where $P$ is the final population in the upper state. It shows that, as expected, hG-RIO achieves much better performance. The relative lack of robustness of the adiabatic process can be explained by the relatively low value of the pulse area $5.84$ (less than $2\pi$) imposed by the RIO protocol.

We recover a comparable robustness (on the left part of the profile) using hG-TCAP of pulse area more than 50\% larger (with the same hyper-Gaussian shape), as shown in Fig. \ref{fig5}, lower frame.

\section{Conclusion and Discussion}

We have analyzed fast robust control, very close to the time-optimal one, using hyper-Gaussian pulses as smooth quasi-square pulses replacing the ideal square pulse. Complete population transfer has been targeted. We have used two protocols: (i) the RIO protocol implementing a smooth pulse, and (ii) TCAP. In TCAP, we have first used an expansion combining a linear term with sine components with free coefficients. We have numerically shown that the time-contraction with an optimization of the free coefficients minimizing its peak induces the pulse to approach a square pulse. We have alternatively designed the contraction to precisely shape a hyper-Gaussian pulse.
We have compared the two protocols using the same hyper-Gaussian control pulse, namely hG-RIO and hG-TCAP. The dynamics in the two cases is very different: in hG-RIO, the dynamics takes much time and oscillate near the half superposition (see Fig. \ref{fig2}, upper frame), which is associated to the multi-valued function $\gamma$ in function of $\theta$ around $\pi/2$ in Fig. \ref{figTraject}. 
In hG-TCAP, the dynamics monotonically reaches the transfer with quasi-linear evolution (except near the beginning and the end, see Fig. \ref{fig4}, upper frame).

We have numerically shown that, as expected, the robustness profile is much better for hG-RIO since this method is precisely constructed to produce an optimal robust dynamics and that the resulting pulse area is relatively too small to achieve an efficient adiabatic passage. However, hG-TCAP is of interest as it allows (i) a fast adiabatic passage (for a given pulse area) and (ii) a profile of robustness comparable to that of RIO using a 50\% larger pulse area.

The main drawback of adiabatic passage in such a two-level system is the lack of control of the final phase of the state, since it is given by the (dynamical) time-integral of the eigenvalue associated to the eigenstate involved in the passage, which is sensitive to the pulse amplitude. This is an important issue to produce quantum gate.
On the other hand, the RIO technique can be easily adapted to feature robustness including the phase of the state \cite{Dridi,Xavier}. 

We highlight that, in addition of being optimal, the RIO protocol features a very simple (oscillatory) shape for the detuning. However, this process, as constructed, is not robust with respect to the detuning; it should be thus preferred when the frequencies are well controlled. Otherwise, the technique has to be extended to be resistant to inhomogeneous broadening [via the parameter $\delta$ in Eq. \eqref{Halpdelt}] or adiabatic passage can be considered if the internal phase is unimportant.

The model presents several practical limitations that are discussed below.

The use of short pulses might involve more than two levels. This would be typically the case in a superconducting qubits system that features a multi-level ladder of relatively weak anharmonicity \cite{superconductor}. In this situation, one can take the additional excited states into account at the lowest order, using perturbation theory. This would add dynamical Stark shift which can be easily treated and compensated by an appropriate detuning design, as a generalization of Ref. \cite{prl103}. Short pulses might also involve a coupling term beyond the RWA. It can be treated via Floquet theory, which, at the lowest order, will also give rise to a dynamical Stark shift, see, e.g., in \cite{parallel1}.

If the system is subjected to noise, such as decoherence due to environmental effects (e.g. dephasing noise or spontaneous emission), one can analyze the following typical situations. If the noise affects permanently the upper state (spontaneous emission), then one has to drive the system in a time much shorter than the typical decoherence time to avoid losses. A typical situation is when the loss drives the dynamics outside the two-state system, it is effectively modeled by a negative imaginary term on the diagonal of the Hamiltonian. This corresponds to resonance coalescence (exceptional points) with their specific adiabatic conditions \cite{EP}. The determination of a robust process could be determined for a given total loss. This has been shown in a different situation with a $\Lambda$-system, featuring a lossy upper state \cite{Xavier}, where robust pulse shapes have been derived for a given total loss. A slow stochastic noise in the energy levels of the qubit (corresponding to a systematic frequency errors  in a quasi-static representation) or a quasi-static transverse noise is represented by the static parameter $\delta$ and $\beta$, respectively, in the model \eqref{Halpdelt}. This will lead to different types of robustness integrals which have to be considered as new constraints in the Euler-Lagrange formulation. 
A fast noise can be modeled by a Lindblad equation (in the Markov approximation) and to an effective Schr\"odinger equation in the case of pure dephasing \cite{pra75dephasing}. The Euler-Lagrange formulation can be applied to such linear (lossy) systems. It has been shown in \cite{whitenoise} that the flat $\pi$-pulse is the least sensitive protocol to phase noise since it is the fastest process. Allowing a given loss would allow here again the derivation of robust protocols in the same way as in \cite{Xavier}.

The implementation of hyper-Gaussian smooth quasi square pulses depends on the timescale considered. In the nano- and sub nano-second regimes, such waveforms can be produced directly in the time-domain by electronics. Below this regime, typically in the femto- and pico-second timescale, the principle of ultrashort pulse shaping is based on the spectral and amplitude modulation of the pulse spectral components \cite{Weiner}, including the liquid crystal programmable spatial light modulator, the acoustic-optic modulator programmable spatial light modulator and the acousto-optic programmable dispersive filter. Such techniques can been used to produce quasi square pulses \cite{Cicaldi}.

\section*{Acknowledgement}
This work was supported by the EUR-EIPHI Graduate School (17-EURE-0002). We also acknowledge support from the European Union's Horizon 2020 research and innovation program under the Marie Sklodowska-Curie grant agreement No. 765075 (LIMQUET). \XC{XC acknowledges the Ramon y Cajal program (Grant No. RYC-2017-22482).
JZ acknowledges additional support from the
CSC (China Scholarship Council).\\
This article is dedicated to the memory of Bruce W. Shore, who has been a great inspiration for SG for so many years, in particular concerning adiabatic and robust quantum control addressed in the present paper.}


\end{document}